\documentclass{jetpl}

\usepackage{bm}
\usepackage{url}
\usepackage{tikz}
\usepackage{cite}

\twocolumn

\lat

\title{The Gailitis-Damburg oscillations in the three-body $e^-e^+\bar{p}$ system}

\rtitle{Gailitis-Damburg oscillations in the three-body $e^-e^+\bar{p}$ system}

\sodtitle{Gailitis-Damburg oscillations in the three-body $e^-e^+\bar{p}$ system}

\author{V.\,A. Gradusov\/\thanks{e-mail: v.gradusov@spbu.ru}, S.\,L. Yakovlev\/\thanks{e-mail: s.yakovlev@spbu.ru}}

\rauthor{V.\,A. Gradusov, S.\,L. Yakovlev}

\sodauthor{Gradusov, Yakovlev}

\address{St. Petersburg State University, St. Petersburg, 199034 Russia}

\dates{\today}{\today}

\abstract{We study the near threshold behavior of cross sections of low-energy antiproton scattering off the ground and excited states of positronium with zero total orbital momentum $L=0$. In our computational experiment, the existence of singularities called the Gailitis-Damburg oscillations above the thresholds of excited states of positronium and antihydrogen atoms is confirmed. In the future the obtained results can be useful for developing proposals for improving the conditions of experiments with antimatter.}

\PACS{03.65.Nk, 34.80.-i}

\begin{document}

\maketitle

Several experiments on antimatter based on the use of the Antiproton Decelerator Facility are being planned and performed at CERN.
Two of them aimed at the antimatter gravitational behavior --- AEgIS~\cite{AEGIS15} and GBAR~\cite{GBAR15} --- use,
inter alia, the three-body reaction
\begin{equation}
\label{react}
\bar{p}+\mathrm{Ps}\to\overline{\mathrm{H}}+e^-
\end{equation}
of antihydrogen $\overline{\mathrm{H}}$ formation via antiproton $\bar{p}$ scattering off the gas of Rydberg positronium (Ps).
In this regard, a number of theoretical works which study the reaction~(\ref{react}) have recently appeared.
Of natural interest here is to find a mechanism for increasing the reaction rate of antihydrogen formation process used for the production of antimatter atoms.

The researchers are interested in various singularities of the cross sections of scattering processes in the $e^-e^+\bar{p}$ system.
Among them are resonances, the near threshold enhancement of cross sections, as well as the less known above threshold singularities called the Gailitis-Damburg (GD) oscillations, predicted for the first time in 
\cite{Gaili63, Gaili63b}. The latter arise due to long-range dipole interaction between an excited atom
(either $\overline{\mathrm{H}}$ or Ps) and a charged particle ($e^-$ or $\bar{p}$).
The GD theory~\cite{Gaili63, Gaili63b, Gaili82} predicts two types of singularities: a series of narrow resonances in energy regions below the thresholds of excited states of atoms and cross-section oscillations above these thresholds.
The existence of the first of them, also called Feshbach resonances, in the $e^-e^-p$ and $e^-e^+\bar{p}$ systems has been reliably confirmed both experimentally
and theoretically, by using very accurate special methods for calculating the energies and widths of resonances~\cite{Bart06,abrash91,chakra07,Yar17,ho04, varga08,yu12,umair14,Burke}.
The situation is more complicated with singularities of another type—oscillations of cross sections.
Their existence in the $e^-e^+\bar{p}$ system has been discussed in the works~\cite{Hu02b, Lazau18, Fabrikant16}, but only in the last of them the cross sections that are consistent with the predictions of the GD theory have been obtained.

The goal of this work is
to study the above threshold behavior of scattering cross sections in the $e^-e^+\bar{p}$ system in the case of zero total orbital momentum of the system $L=0$.
Our ab initio approach to solving the multichannel quantum scattering problem in a system of three particles is based on the solution of the Faddeev-Merkuriev (FM) equations, which are strictly equivalent to the Schrödinger equation~\cite{FM}, in the configuration space.
These equations in the total orbital momentum representation~\cite{Kostr89} are reduced to a finite set of three-dimensional partial differential equations.
To solve boundary value problems for these equations, we proposed and implemented an efficient computational algorithm~\cite{Grad21}, which was tested, in particular, in calculations of low-energy scattering in the $e^-e^+\bar{p}$ system~\cite{Grad21b}.
To calculate the obtained in this work cross sections above the thresholds of excited states of atoms, it is critically important to use the asymptote of solutions to the FM equations, which explicitly takes into account the long-range nature of effective interaction between a neutral atom and an incident (outgoing) particle.
By this reason one needs to modify the ``standard'' formulae for the asymptotic behavior of wave functions~\cite{Gaili76, Burke}.
This modification is a generalization to the three-particle case of the results of our work \cite{Yak23b}.
Here we briefly discuss the corresponding theory and apply it in a series of calculations of low-energy cross sections above the thresholds of the first few excited states of atoms in the $e^-e^+\bar{p}$ system.

A system of three spinless nonrelativistic charged particles with masses $m_{\alpha}$ and charges $Z_\alpha$, $\alpha=1,2,3$ is considered.
In what follows, the set of indices $\alpha\beta\gamma$ denotes one of the cyclic permutations of the set $\{1,2,3\}$ that enumerates the particles.
Since in the triad $\alpha\beta\gamma$ the pair of particles $\beta\gamma$ is completely determined by the number $\alpha$ of the third particle, we will systematically use this fact to identify pairs of particles.
In the center of mass frame, the position of particles is described by a set of Jacobi coordinates.
For the partition $\alpha(\beta\gamma)$, they are defined as the relative position vectors between $\beta\gamma$ particles $\bm{x}_\alpha$
and between their center of mass and the particle $\alpha$ $\bm{y_\alpha}$.
We use reduced Jacobi coordinates $\{\bm{x}_\alpha, \bm{y}_\alpha\}$, which are Jacobi vectors scaled by factors $\sqrt{2\mu_{\alpha}}$ and $\sqrt{2\mu_{\alpha(\beta\gamma)}}$ respectively.
Reduced masses of the pair $\alpha$ ($\mu_{\alpha}$) and the system ``particle $\alpha$ --- pair $\alpha$'' ($\mu_{\alpha(\beta\gamma)} $) are expressed in terms of particle masses $m_\alpha$ by standard formulae.
For different values of $\alpha$, the reduced Jacobi vectors are related by an orthogonal transform $\bm{x}_\beta=c_{\beta\alpha}\bm{x}_\alpha + s_{\beta\alpha}\bm{y }_\alpha$, $\bm{y}_\beta=-s_{\beta\alpha}\bm{x}_\alpha + c_{\beta\alpha}\bm{y}_\alpha$~\cite{FM}. In what follows, the vector lengths are denoted by the corresponding non-bold letters, for example, $x_\alpha=|\bm{x}_\alpha|$.
The states of a system with total orbital momentum $L=0$ are symmetric with respect to the rotation of the system as a whole and for this reason depend only on the three coordinates $X_\alpha = \{x_\alpha,y_\alpha,z_\alpha=\cos\theta_ \alpha\equiv(\bm{x}_\alpha,\bm{y}_\alpha)/(x_\alpha y_\alpha)\}$ that determine the position of particles in the plane containing them.
In what follows, where it is due, it is assumed that the coordinates $X_\beta$ are expressed in terms of $X_\alpha$.

The FM equations for three charged particles~\cite{FM, merkur80} in the case of $L=0$ have the form~\cite{Grad21b, Grad19}:
\begin{multline}
\label{FMeq}
\left[ T_\alpha + V_\alpha(x_\alpha)+\sum_{\beta\ne\alpha}V_\beta^{(\text{l})}(x_\beta,y_\beta) - E \right]\psi_\alpha(X_\alpha) = \\
- V_\alpha^{(\text{s})}(x_\alpha,y_\alpha)\sum_{\beta\ne\alpha}\frac{x_\alpha y_\alpha}{x_\beta y_\beta }\psi_\beta(X_\beta).
\end{multline}
Here the kinetic energy operators are given by
\begin{equation}
T_\alpha = -\frac{\partial^2}{\partial y_\alpha^2} -\frac{\partial^2}{\partial x_\alpha^2} - \left(\frac{1}{ y_\alpha^2}+\frac{1}{x_\alpha^2}\right)\frac{\partial}{\partial z_\alpha}(1-z_\alpha^2)\frac{\partial} {\partial z_\alpha}.
\end{equation}
The potentials $V_\alpha$ represent the pair Coulomb interaction $V_{\alpha}(x_\alpha)=\sqrt{2\mu_{\alpha}}Z_\beta Z_\gamma/x_\alpha$.
They are split into short-range $V^{(\mathrm{s})}_\alpha$ and long-range parts $V^{(\mathrm{l})}_\alpha$
\begin{equation}
\label{PotSplit}
V_\alpha(x_\alpha) = V^{(\mathrm{s})}_\alpha(x_\alpha,y_\alpha) + V^{(\mathrm{l})}_\alpha(x_\alpha, y_\alpha)
\end{equation}
using a function called the Merkuriev cut-off~\cite{FM, Grad21b}.
The equations~(\ref{FMeq}) can be summed up, which leads to the Schrödinger equation for the wave function $\Psi=\sum_{\alpha}\psi_\alpha(X_\alpha)/(x_\alpha y_\alpha)$, where $\psi_{\alpha}$ are the components of the wave function given by the solution of the equations~(\ref{FMeq}).

At energy values $E$ below the threshold of breakup (ionization) of the system, the FM components $\psi_\alpha(X_\alpha)$ at $y_\alpha \to \infty$ are substantially different from zero only
in the asymptotic region $\Omega_\alpha =\left\{ x_\alpha, y_\alpha: \, x_\alpha \ll y_\alpha \right.$ for $\left. y_\alpha\to\infty\right\}$. In $\Omega_\alpha$ the FM components can be represented in the form
\begin{multline}
\label{asympt}
\psi_\alpha^{(\alpha_0\nu\lambda)}(X_\alpha) \thicksim \\
\thicksim \sum_{n\ell}\phi_{n\ell}(x_\alpha)Y_{\ell0}(\theta_\alpha,0)
\bigg( \psi_{(n\ell)(\nu\lambda)}^-(y_\alpha, p_{\nu})\delta_{\alpha\alpha_0}- \\
- \sum_{n'\ell'}\psi_{(n\ell)(n'\ell')}^+(y_\alpha, p_{n'})\sqrt{\frac{p_{\nu} }{p_{n'}}}\mathcal{S}_{(\alpha n'\ell')(\alpha_0\nu\lambda)} \bigg).
\end{multline}
In this formula
the indices $\alpha n\ell$ enumerate the binary scattering channels, i.e., the Coulomb bound states of two particles in pair $\alpha$ with radial wave function $\phi_{n\ell}(x)$ and energy $\varepsilon_n$.
The set of indices $\alpha_0\nu\lambda$ defines the initial scattering channel.
$Y_{\ell m}$ denotes the standard spherical harmonic.
In formula~(\ref{asympt}) and below in the text it is assumed that the indices $n\ell$ take integer values $n>\ell\ge0$ corresponding to channels which are open at a given energy $E$.
The momentum $p_n$ of an incident (scattered) particle is determined by the energy conservation condition $E = p_n^2+\varepsilon_n$. Accordingly, the channel is considered open if
$E-\varepsilon_n\ge 0$.
The functions $\psi_{(n\ell)(\nu\lambda)}^-$ and $\psi_{(n\ell)(\nu\lambda)}^+$ define the incident and scattered waves.
It is standard to choose  these functions in the form
\begin{equation}
\label{stand}
\widehat{\psi}_{(n\ell)(n'\ell')}^{\pm}(y_\alpha, p_{n'}) = u_{\ell}^{\pm}(\eta_n, p_n y_\alpha)\delta_{(n\ell)(n'\ell')},
\end{equation}
where $u_{\ell}^{\pm}(\eta,z)$  are the Coulomb incoming and outgoing waves~\cite{Mess},
and the Sommerfeld parameter is defined as $\eta_n\equiv Z_\alpha(Z_\beta+Z_\gamma)\sqrt{2\mu_{\alpha(\beta\gamma)}}/(2p_n)$.
Indeed, the use of functions~(\ref{stand}) in~(\ref{asympt}) leads to the solution of the FM equations with asymptotic behavior of the form:
\begin{multline}
\label{as-st}
\widehat{\psi}^{(\alpha_0\nu\lambda)}_\alpha(X_\alpha) \thicksim \\
\thicksim \phi_{\nu\lambda}(x_\alpha)Y_{\lambda 0}(\theta_\alpha,0)u_\lambda^-(\eta_{\nu}, p_\nu y_\alpha)\delta_{\alpha\alpha_0}-\\
  -\sum_{n\ell}\phi_{n\ell}(x_\alpha)Y_{\ell0}(\theta_\alpha,0) u_\ell^+(\eta_{n}, p_n y_\alpha)
\sqrt{\frac{p_{\nu}}{p_{n}}}\widehat{\mathcal{S}}_{(\alpha n\ell)(\alpha_0\nu\lambda)}.
\end{multline}
The cross section of the scattering process with initial $\alpha_0\nu\lambda$ and final $\alpha n\ell$ channels is expressed in a standard way through the S-matrix element $\widehat{\mathcal{S}}_{(\alpha n\ell) ,(\alpha_0\nu\lambda)}$~\cite{Grad19}.

In a system of three charged particles, the presence of an effective dipole potential between the excited bound state of pair $\alpha$ (atom) and particle $\alpha$ leads to the fact that the representation~(\ref{stand}) becomes insufficiently accurate, since after substituting it into the FM equations the dipole potential 
is not compensated.
Formally, the dipole potential can be obtained by substituting into the FM equations~(\ref{FMeq}) the asymptotic expansion in $\Omega_\alpha$ of the sum of the long-range parts of the potentials
\begin{equation}
\label{pot-as}
\sum_{\beta\ne\alpha}V_\beta^{(\text{l})}(x_\beta,y_\beta) =
\sum_{\ell=0}^\infty d_\alpha^{(\ell+1)}\frac{x_\alpha^\ell P_\ell(z_\alpha)}{y_\alpha^{\ell+ 1}},
\end{equation}
where $P_\ell$ are Legendre polynomials.
Indeed, from the properties of the Merkuriev cut-off it follows that in $\Omega_\alpha$ the quantity $V_\beta^{(\text{l})}$, up to a term exponentially decreasing in the variable $y_\alpha$, coincides with the potential $V_\beta$. Then the coefficients of the multipole expansion~(\ref{pot-as}) can be obtained by using the formula
\begin{eqnarray}
\frac{1}{x_\beta} & = &
\left.\frac{1}{| c_{\beta\alpha}\bm{x}_\alpha+s_{\beta\alpha}\bm{y}_\alpha |}\right|_{|s_{\beta\alpha}|y_\alpha \ge |c_{\beta\alpha}|x_\alpha} \nonumber \\
& = &
\frac{1}{|s_{\beta\alpha}|y_\alpha}\sum_{\ell=0}^\infty
\left( \frac{|c_{\beta\alpha}|x_\alpha}{s_{\beta\alpha}y_\alpha} \right)^\ell P_\ell(z_\alpha).
\end{eqnarray}
In particular, the first two coefficients have the form
\begin{equation}
C_\alpha \equiv d_\alpha^{(1)} = Z_\alpha(Z_\beta+Z_\gamma)\sqrt{2\mu_{\alpha(\beta\gamma)}},
\end{equation}
\begin{multline}
D_\alpha \equiv d_\alpha^{(2)} =
Z_\alpha(-1)^{\alpha}\sqrt{2\mu_{\alpha(\beta\gamma)}}\sqrt{\frac{m_\alpha}{m_\alpha+m_\beta+m_\gamma}}\\
\times\bigg[
Z_\gamma\,\text{sign}(\beta-\alpha)(-1)^{\beta}\sqrt{\frac{m_\beta}{m_\gamma}} \\
+Z_\beta\,\text{sign}(\gamma-\alpha)(-1)^{\gamma}\sqrt{\frac{m_\gamma}{m_\beta}}
\bigg].
\end{multline}
As is known from the theory of the FM equations, the right-hand sides of the equations decrease exponentially in $\Omega_\alpha$.
Substituting the first two terms of the expansion~(\ref{pot-as}) into the FM equations~(\ref{FMeq}), we obtain that in the asymptotic region $\Omega_\alpha$ the equations
take the form
\begin{multline}
\label{FMeq-as}
\left[ T_\alpha + V_\alpha(x_\alpha) + \frac{C_\alpha}{y_\alpha} + \frac{D_\alpha x_\alpha z_\alpha}{y_\alpha^2} - E\right]\psi_\alpha^{(\alpha_0\nu\lambda)}(X_\alpha)\\
= O\left(\frac{1}{y_\alpha^3}\right).
\end{multline}
  Let us now substitute the asymptotic representation~(\ref{asympt}) into the equations~(\ref{FMeq-as}) and project them onto the bound states wave functions $\phi_{n\ell}Y_{\ell0}$.
We use the following well-known relations~\cite{Mess,NISTDLMF}:
\begin{equation}
\frac{\partial}{\partial \cos\theta}(1-\cos^2\theta)\frac{\partial}{\partial \cos\theta} Y_{\ell0}(\theta,0) =
-\ell(\ell+1)Y_{\ell0}(\theta,0),
\end{equation}
\begin{equation}
\left[ -\frac{d^2}{d x_\alpha^2} + \frac{\ell(\ell+1)}{x_\alpha^2} + V_\alpha(x_\alpha) - \varepsilon_n \right]\phi_{n\ell}(x_\alpha) = 0,
\end{equation}
\begin{multline}
2\pi\int_0^{+\infty}\text{d}x_\alpha\int_{-1}^{1}\text{d}\cos\theta\,\phi_{n'\ell'}( x_\alpha)Y_{\ell'0}(\theta,0)\phi_{n\ell}(x_\alpha)Y_{\ell0}(\theta,0) \\
= \delta_{\ell\ell'}\delta_{nn'},
\end{multline}
\begin{multline}
2\pi\int_{-1}^{1}\text{d}\cos\theta\,Y_{\ell'0}(\theta,0)\cos\theta Y_{\ell0}(\theta, 0)\\
= \delta_{\ell,\ell'+1}\frac{\ell}{\sqrt{4\ell^2-1}}+\delta_{\ell,\ell'-1}\frac{\ell +1}{\sqrt{4(\ell+1)^2-1}}.
\end{multline}
As a result, we obtain that the functions $\psi_{(n\ell)(\nu\lambda)}^{\pm}$ are the linearly independent solutions to the close coupling equations:
\begin{multline}
\label{CC}
\left[
-\frac{d^2}{d y_\alpha^2} + \frac{\ell(\ell+1)}{y_\alpha^2} + \frac{C_\alpha}{y_\alpha} - p_n^2
\right]
\psi_{(n\ell)(\nu\lambda)}^{\pm}(y_\alpha, p_{\nu}) \\
+\sum_{n'\ell'}\frac{A^\alpha_{n\ell,n'\ell'}}{y_\alpha^2}\psi_{(n'\ell')(\nu\lambda)}^{\pm}(y_\alpha, p_{\nu}) = O\left(\frac{1}{y_\alpha^3}\right).
\end{multline}
The elements of the matrix $A^{\alpha}$, which specifies the effective dipole potential, are given by the expressions
\begin{eqnarray}
&A^\alpha_{n\ell,n'\ell'} =
D_\alpha M^\alpha_{n\ell,n'\ell'}\times \nonumber \\
& \times \bigg( \delta_{\ell',\ell+1}\frac{\ell+1}{\sqrt{4(\ell+1)^2-1}}
+ \delta_{\ell',\ell-1}\frac{\ell}{\sqrt{4\ell^2-1}} \bigg),
\end{eqnarray}
where
\begin{equation}
M^\alpha_{n\ell,n'\ell'} \equiv \int_0^{+\infty}\text{d}x_\alpha \phi_{n'\ell'}(x_\alpha)x_\alpha \phi_{n\ell}(x_\alpha).
\end{equation}

As is mentioned above, the incoming and outgoing waves $\widehat{\psi}_{(n\ell)(n'\ell')}^{\pm}$ defined in~(\ref{stand}) do not accurately enough describe the asymptotic behavior of the solution to the FM equations in $\Omega_\alpha$, since they do not take into account the presence of an effective dipole potential.
Indeed, when substituting these functions into equations (\ref{CC}), the last dipole term of order $ O(y^{-2}_\alpha)$ remains uncanceled on the left side.
The dipole term in the asymptotic solutions of the equations (\ref{CC}) was partially accounted for in the works \cite{Gaili63, Gaili63b, Gaili76, Burke} by diagonalizing the
block of the channel coupling matrix
\begin{align}
\delta_{nn'} [\ell(\ell+1)\delta_{\ell\ell'}+A^\alpha_{(n\ell)(n'\ell')}], \nonumber \\
\ell= 0.1,\ldots,n-1, \ \ \ell'= 0.1,\ldots,n'-1 \nonumber.
\end{align}
However, in this case, the non-diagonal with respect to $n$ part of the dipole interaction remains uncanceled.
We have taken the full account of the dipole part of the interactions in the equations (\ref{CC}) using direct asymptotic methods, which have led us to the following form
of solutions
\begin{multline}
\label{CC-fun}
\psi_{(n\ell)(\nu\lambda)}^{\pm}(y_\alpha,p_\nu)= \\
\left[
W^{\alpha(0)}_{(n\ell)(\nu\lambda)}+\frac{1}{y_\alpha^2}W^{\alpha(1)}_{(n\ell)(\nu\lambda)}
\right]
u_{L_\alpha^{(\nu\lambda)}}^\pm(\eta_\nu,p_\nu y_\alpha).
\end{multline}
Here the matrices $W^{\alpha(0)}$ and $W^{\alpha(1)}$ are given by the formulae
\begin{eqnarray}
W^{\alpha(0)}_{(n\ell)(\nu\lambda)} & = & \delta_{n\nu}V_{\ell}^{\alpha(\nu\lambda)}, \nonumber\\
W^{\alpha(1)}_{(n\ell)(\nu\lambda)} & = & (1-\delta_{n\nu})\frac{\sum_{\ell'=0}^ {\nu-1}A^\alpha_{(n\ell)(\nu\ell')}V_{\ell'}^{\alpha(\nu\lambda)}}{(p_n^2-p_\nu^2)},
\end{eqnarray}
and new values of orbital momentum $L_\alpha^{(\nu\lambda)}$ are the solutions to the quadratic equation
\begin{equation}
L_\alpha^{(\nu\lambda)}(L_\alpha^{(\nu\lambda)}+1) = q_\alpha^{(\nu\lambda)}.
\end{equation}
Finally, $q_\alpha^{(\nu\lambda)}$, $V^{\alpha(\nu\lambda)}$ are the eigenvalues and eigenvectors of the matrix
\begin{equation}
\ell(\ell+1)\delta_{\ell\ell'}+A^\alpha_{(\nu\ell)(\nu\ell')},\quad
\ell,\ell' = 0.1,\ldots,\nu-1.
\end{equation}
A detailed derivation of the above asymptotic solutions is beyond the scope of this work and will be made in a separate publication. 

The solutions (\ref{CC-fun}) allow us to reformulate the asymptotic boundary conditions (\ref{as-st}) for the FM equations
\begin{multline}
\label{asympt-fin}
\widetilde{\psi}_\alpha^{(\alpha_0\nu\lambda)}(X_\alpha) \thicksim\\
\thicksim \sum_{n\ell}\phi_{n\ell}(x_\alpha)Y_{\ell0}(\theta_\alpha,0)
\times \bigg[ \widetilde{\psi}_{(n\ell)(\nu\lambda)}^-(y_\alpha, p_{\nu})\delta_{\alpha\alpha_0}- \\
- \sum_{n'\ell'}\psi_{(n\ell)(n'\ell')}^+(y_\alpha, p_{n'})\sqrt{\frac{p_{\nu} }{p_{n'}}}\widetilde{\mathcal{S}}_{(\alpha n'\ell')(\alpha_0\nu\lambda)} \bigg],
\end{multline}
with an incident wave of the form
\begin{multline}
\left[\widetilde{\psi}^-\right]_{(n\ell)(\nu\lambda)}(y_\alpha,p_\nu) \\ = \sum_{\lambda'}e^ {i\left(\lambda-L_\alpha^{(\nu\lambda')}\right)\pi/2}\left[V_{\lambda}^{\alpha(\nu\lambda')}\right]^*\psi_{(n\ell)(\nu\lambda')}^-(y_\alpha,p_\nu).
\end{multline}
An important fact is that the right-hand sides (\ref{as-st}) and (\ref{asympt-fin}) coincide in the limit $y_\alpha \to \infty$.
Then it follows that the connection between the components of the ``physical'' S-matrix $\widehat{\mathcal{S}}$ and the matrix $\widetilde{\mathcal{S}}$ defined by the solution~(\ref{asympt-fin}) is given by the equality
\begin{multline}
\label{S-rec}
\widehat{\mathcal{S}}_{(\alpha n\ell)(\alpha_0\nu\lambda)} \\
= \sum_{\ell'}e^{i\left(\ell-L_\alpha^{(n\ell')}\right)\pi/2}V_\ell^{\alpha(n\ell' )}\widetilde{\mathcal{S}}_{(\alpha n\ell')(\alpha_0\nu\lambda)}.
\end{multline}
To solve the FM equations with asymptotic boundary conditions (\ref{asympt-fin}), we use a numerical scheme, described in detail in~\cite{Grad21, PCT23}.
The use of more accurate asymptote~(\ref{asympt-fin}) in calculations leads to a significant reduction in the requirements for computer resources.
This is due to the fact that this asymptote is reached by the FM components at distances significantly smaller than the asymptote~(\ref{as-st}), the latter enforcing to use the size of the computational domain for the variable $y_\alpha$ in hundreds of atomic units~\cite{ Grad23b}.
When turning to sufficiently small above threshold energies $p_n^2$, which we are interested in, this size grows unlimitedly, which makes the calculation of scattering cross sections at such energies almost impossible.
In our work~\cite{Yak23b} we have demonstrated this on the example of a model problem of single-channel scattering off a dipole central potential.

To obtain the presented in the article results, we have calculated the scattering cross sections with an accuracy of no worse than 1\% and a high energy resolution: $6\cdot 10^{-6}$ a.u. when calculating cross sections directly above the thresholds of excited states of atoms and $6\cdot 10^{-5}$ a.u. in other cases.
All presented values are given in atomic units, cross sections are given
in units of $\pi a_0^2$.
Binary scattering processes are specified by the initial and final atomic states. For example, ${\mathrm{Ps}(1)\to\overline{\mathrm{H}}(2)}$ denotes an excited $n=2$ (both $s$ and $p$ states) antihydrogen formation process when antiproton is scattering off the ground ($n=1$) state of positronium.

According to the GD theory~\cite{Gaili82}, the near-threshold oscillations in cross sections arise when some of the new values of orbital momentum $L_\alpha^{(n\ell)}$ are non real.
Above the threshold of the excited bound state of an atom with principal quantum number $n$, in the case of a single (among values with different $\ell<n$) non real value $L_\alpha^{(n\ell)}$, the theory predicts the following dependence of the cross sections on energies $p_n^2$:
\begin{equation}
\label{GD}
\sigma =
A + B\cos(2\Im\mbox{m} \,L_\alpha^{(n\ell)}\ln p_n+\phi).
\end{equation}
Here the constants $A$, $B$, $\phi$, their own for each specific system and section, can be considered to be independent of the energy $p_n^2$ for small $p_n$.
A simple calculation shows that in the system $e^-e^+\bar{p}$ for the first few scattering channels of excited states Ps(2), $\overline{\mathrm{H}}$(3) and $\overline{\mathrm{H}}$(4) the condition described above is realized.
The imaginary parts of the momentum $\Im\mbox{m}\, L_\alpha^{(n\ell)}$ are equal to 4.76914, 2.19836 and 3.99364, respectively.
Thus, in cross sections above the thresholds of these states one can expect the presence of GD oscillations.

\begin{figure*}[t]
\begin{minipage}[h]{0.49\linewidth}
\center{\includegraphics[width=0.99\textwidth]{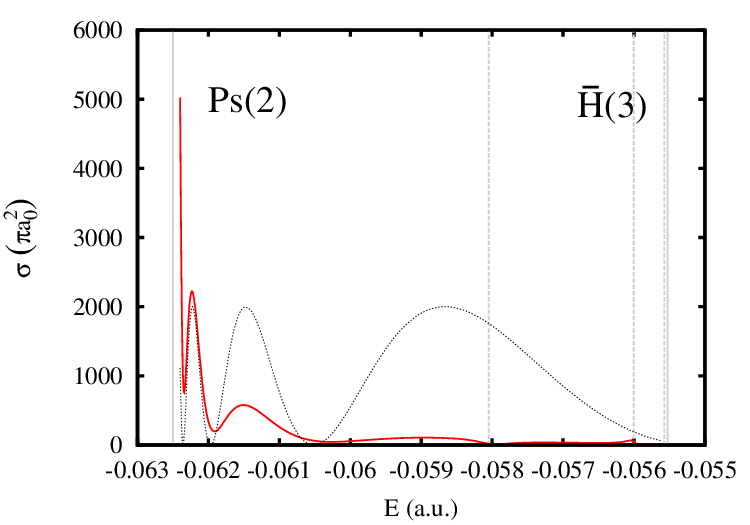}}\\
(a)
\end{minipage}
\hfill
\begin{minipage}[h]{0.49\linewidth}
\center{\includegraphics[width=0.99\textwidth]{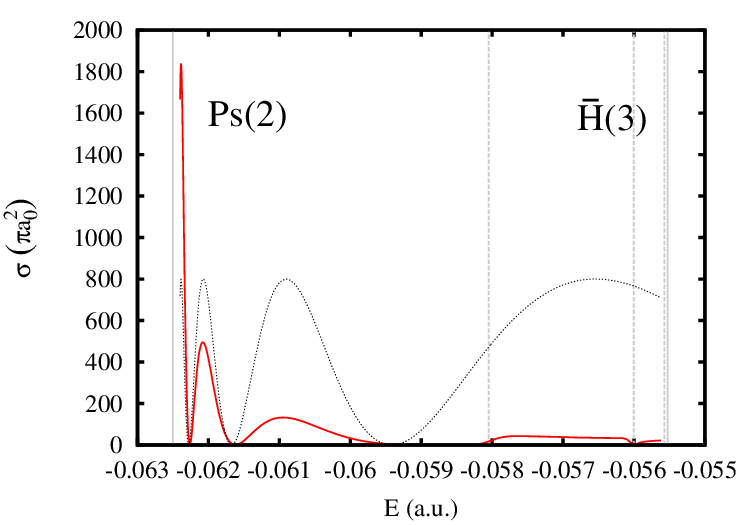}}\\
(b)
\end{minipage}
\caption{Fig.1. Cross sections of (a) elastic Ps(2s)$\to$Ps(2s) and (b) quasi-elastic Ps(2s)$\to$Ps(2p) scattering. The dotted line shows a possible form of the curve~(\ref{GD}). Vertical dashed lines show the positions of the resonances~\cite{ho04, varga08, yu12, umair14}.}
\label{Ps2}
\end{figure*}

Figure~\ref{Ps2} shows the elastic and quasi-elastic cross sections for antiproton scattering off positronium Ps in the first excited state between the thresholds Ps(2) and $\overline{\mathrm{H}}$(3).
In cross sections above the Ps(2) threshold, GD oscillations are clearly visible with the location of maxima being in good agreement with the law~(\ref{GD}).
Indeed, for clarity, Fig.~\ref{Ps2} also shows a graph of the function~(\ref{GD}) with empirically chosen values of constants $A$, $B$ and phase $\phi$.
The presented curves confirm the previously obtained results~\cite{Fabrikant16}, where the feasibility of the law~(\ref{GD}) was verified on the same cross sections.
The first wave of the oscillations presented in Fig.~\ref{Ps2} has also been obtained in the works~\cite{Hu02b,Lazau18}.

\begin{figure}
\includegraphics[width=0.5\textwidth]{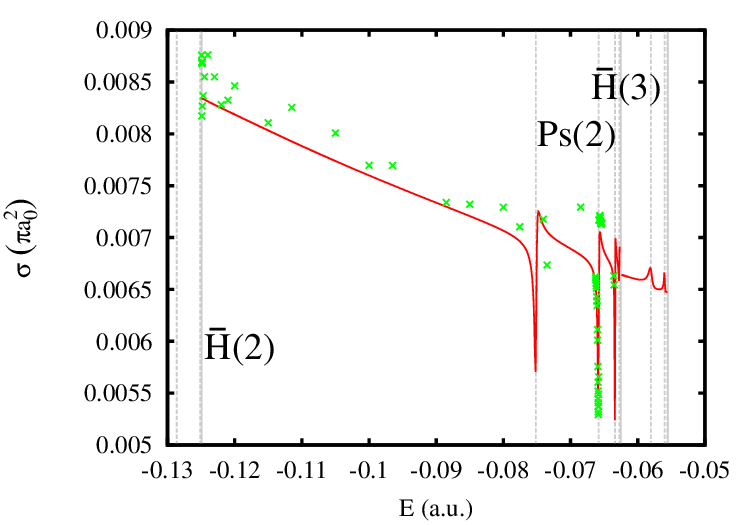}
\caption{Fig.2. Antihydrogen formation cross sections Ps(1)$\to\overline{\mathrm{H}}$(1). Crosses mark points corresponding to the work~\cite{Lazau18} (received from Dr. R. Lazauskas in private communication). Vertical dashed lines show the positions of the resonances~\cite{ho04, varga08, yu12, umair14}.}
\label{Ps1H1}
\end{figure}

\begin{figure}
\includegraphics[width=0.5\textwidth]{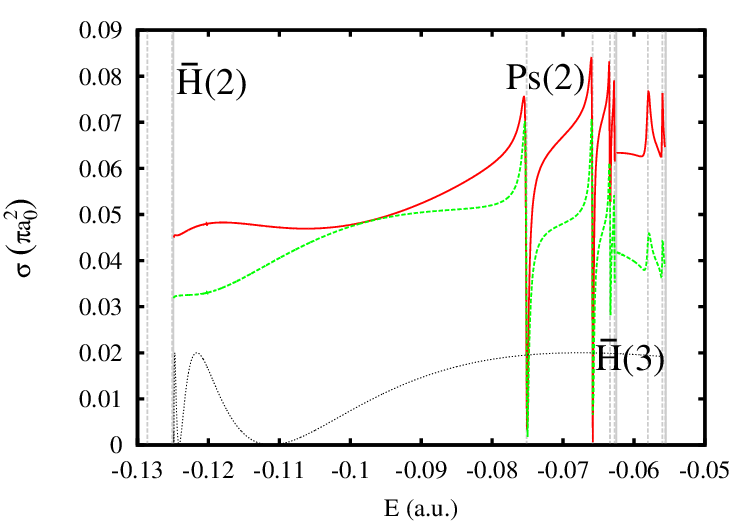}
\caption{Fig.3. Antihydrogen formation cross sections Ps(1)$\to\overline{\mathrm{H}}$(2s) (solid) and Ps(1)$\to\overline{\mathrm{H}}$(2p) (dashed line). The dotted line shows a possible form of the curve~(\ref{GD}). Vertical dashed lines show the positions of the resonances~\cite{ho04, varga08, yu12, umair14}.}
\label{Ps1H2}
\end{figure}

Let us now move on to a discussion of the antihydrogen $\overline{\mathrm{H}}$ formation cross sections above various thresholds of excited states of atoms.
Figures~\ref{Ps1H1} and~\ref{Ps1H2} show the cross sections of the formation of antihydrogen $\overline{\mathrm{H}}$ in the ground  $\overline{\mathrm{H}}$(1) and excited $ \overline{\mathrm{H}}$(2s), $\overline{\mathrm{H}}$(2p) states between the $\overline{\mathrm{H}}$(2) and $\overline {\mathrm{H}}$(3) states thresholds.
This is a refinement of our previously published results~\cite{Grad19}, which have been obtained using standard asymptotic boundary conditions~(\ref{as-st}). This boundary conditions have had to be imposed at very large intercluster distances $y_\alpha$ to achieve convergence of results of calculations.
The latter has required the involvement of very serious computer resources, however, even this in some cases has not allowed us to achieve the required accuracy at low above-threshold energies.
In addition, in our previous work~\cite{Grad21b} in Fig.~4 the summed cross section Ps(2)$\to\overline{\mathrm{H}}$(1,2) in the energy interval between the thresholds Ps(2) and $\overline{\mathrm{H}}$(3) is shown.
We do not duplicate it here to save space.
Among all the mentioned cross sections, weak oscillations can be seen only in Fig.~\ref{Ps1H2} in the cross sections of the formation of antihydrogen in excited states $\overline{\mathrm{H}}$(2) above the corresponding threshold.
Due to small amplitudes of these oscillations and small number of waves, it is difficult to definitely conclude whether they are related to the threshold behavior predicted by GD theory.
However, the existing oscillations are consistent with the law~(\ref{GD}), which is illustrated in Fig.~\ref{Ps1H2}.
All the cross sections have a fairly smooth character everywhere, except in the vicinity of the below threshold resonances marked in the figures.
In particular, we do not see in our results the obtained in the works~\cite{Hu02b,Lazau18} sharp peaks in the cross sections Ps(1,2)$\to\overline{\mathrm{H}}$(1,2) just above the threshold of the excited Ps(2) state.

\begin{figure}
\includegraphics[width=0.5\textwidth]{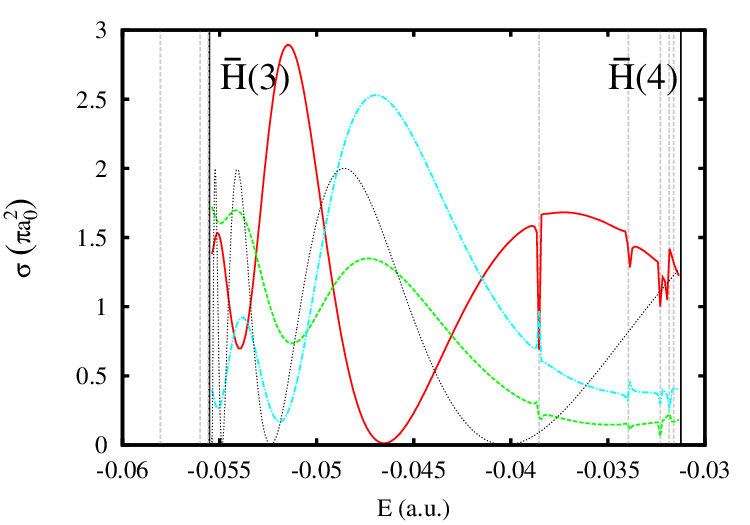}
\caption{Fig.4. Antihydrogen formation cross sections Ps(2s)$\to\overline{\mathrm{H}}$(3s) (solid), Ps(2s)$\to\overline{\mathrm{H}}$(3p) (dashed) and Ps(2s)$\to\overline{\mathrm{H}}$(3d) (dash-dotted line). The dotted line shows a possible form of the curve~(\ref{GD}). Vertical dashed lines show the positions of the resonances~\cite{ho04, varga08, yu12, umair14}.}
\label{Ps2H3}
\end{figure}

Finally, in the cross sections of the formation of antihydrogen in the second excited state $\overline{\mathrm{H}}$(3), shown in Fig.~\ref{Ps2H3}, we have discovered oscillations, the positions of maxima of which satisfy the dependence~(\ref{GD}) in general.
Note that the GD theory predicts small relative amplitudes of oscillations in the cross sections of transition processes from old channels to new channels that emerge above the corresponding threshold~\cite{Gaili82}.
This statement, generally speaking, does not agree with the form of cross section oscillations in Fig.~\ref{Ps2H3}, since the latter have fairly large amplitudes.
Perhaps this circumstance is due to the fact that the GD theory, as stated above, does not take into account the dipole interaction between channels
with different values of $n$.
One challenge for future research may be to further identify the reasons for this inconsistency.
We also plan to generalize the theory of taking into account dipole interaction in the case of $L>0$ and carry out corresponding high-precision calculations of scattering cross sections in the $e^-e^+\bar{p}$ system.
We hope that this will make it possible to more definitely answer the question about the existence of GD oscillations in the total cross sections of scattering processes which are directly measured in experiment.

------------------

We acknowledge the support of the Russian Science Foundation, project no. 23-22-00109.
Research was carried out using computational resources
provided by Resource Center ``Computer Center of SPbU''
(http://cc.spbu.ru).
The authors express their thanks to V.A. Roudnev и E.A. Yarevsky for fruitful discussions of the results of this work. 


\end{document}